\documentclass[nofootinbib,twocolumn,superscriptaddress]{revtex4}
\usepackage{amsmath}
\usepackage{amsthm}
\usepackage{mathtools}
\usepackage{amssymb}
\usepackage{dsfont}
\usepackage{subfigure}
\usepackage{upgreek}
\usepackage[english]{babel}
\begin{document}

\title{Experimental investigation of the uncertainty relations with coherent light}

\author{Hui Wang}
\affiliation{Center of Materials Science and Optoelectronics Engineering \& CMSOT, \\ University of Chinese Academy of Sciences, YuQuan Road 19A, Beijing 100049, China}

\author{Jun-Li Li}
\affiliation{Center of Materials Science and Optoelectronics Engineering \& CMSOT, \\ University of Chinese Academy of Sciences, YuQuan Road 19A, Beijing 100049, China}

\author{Shuang Wang}
\affiliation{Center of Materials Science and Optoelectronics Engineering \& CMSOT, \\ University of Chinese Academy of Sciences, YuQuan Road 19A, Beijing 100049, China}

\author{Qiu-Cheng Song}
\affiliation{School of Physical Sciences \& Key Laboratory of Vacuum Physics, \\ University of Chinese Academy of Sciences, YuQuan Road 19A, Beijing 100049, China}

\author{Cong-Feng Qiao}
\email{qiaocf@ucas.ac.cn}
\affiliation{School of Physical Sciences \& Key Laboratory of Vacuum Physics, \\ University of Chinese Academy of Sciences, YuQuan Road 19A, Beijing 100049, China}

\begin{abstract}
  Taking advantage of coherent light beams, we experimentally investigate the variancebased  uncertainty relations and the optimal majorization uncertainty relation for the  two-dimensional quantum mechanical system. Different from most of the experiments which devoted to record each individual quantum, we examine the uncertainty relations by measuring an ensemble of photons with two polarization degree of freedom characterized by the Stokes parameters which allow us to determine the polarization density matrix with high precision. The optimality of the recently proposed direct-sum majorization uncertainty relation is verified by measuring the Lorenz curves. Results show that the Lorenz curve method represents a faithful verification of the majorization uncertainty relation and the uncertainty relation is indeed an ensemble property of quantum system.
\end{abstract}
\maketitle

\section{introduction}
Uncertainty relation is one of the symbolic features of quantum mechanics, and behaves as a fundamental limitation on the precision measurements of incompatible observables. Since the seminal work of Heisenberg \cite{heis}, investigations by Kennard \cite{Kennard}, Weyl \cite{Weyl}, and Robertson \cite{Robertson} give the rigorous derivation of the uncertainty relations, among which there is the following famous formula
\begin{equation}
\Delta X^2\Delta Y^2\geq
\frac{1}{4}|  \langle [X,Y] \rangle |^2 \; .  \label{Robertson1}
\end{equation}
Here, $\Delta X^2$ represents the variance of the observable $X$ and the commutator is $[X,Y]:=XY-YX$. The inequality (\ref{Robertson1}) predicts a trade-off relation between variances of two observables, lower bounded by the expectation value of their commutator. Except for the canonical commutation relation $[x,p]=i\hbar$, the expectation of the commutator may be zero and the uncertainty relation (\ref{Robertson1}) then turns out to be trivial.

To avoid state dependence of the lower bounds, entropic uncertainty relations were developed \cite{Entropic-Un-1}. A typical form of the entropic uncertainty relation reads \cite{Entropic-Un-2}
\begin{equation}
H(\vec{p}_X) +H(\vec{p}_Y) \geq \log \frac{1}{c} \; , \label{entro-c}
\end{equation}
where $H(\cdot)$ is the Shannon entropy of the probability distribution $\vec{p}_{X,Y}$ for the measurement outcomes of the observable $X$($Y$) and $c:=\max_{i,j}|\langle x_i|y_j\rangle|^2$ is the maximum overlap of the eigenvectors of $X$ and $Y$. Similar uncertainty relations also hold for other entropic functions, e.g. R\'enyi entropies \cite{Entropic-funs}, and the corresponding lower bounds vary with the chosen functions. For this reason, the ``universal uncertainty relation'' was proposed \cite{Maj-1}
\begin{equation}
\vec{p}_X \otimes \vec{p}_Y \prec \vec{\omega} \; . \label{Maj-un}
\end{equation}
Here, $\otimes$ is the direct product and $\vec{\omega}$ is a state independent vector. The majorization relation between two $N$-dimensional vectors $\vec{p} \prec \vec{q}$ means $\forall k\in \{1,\cdots, N\}$, $\sum_{\mu=1}^k p^{\downarrow}_{\mu} \leq \sum_{\nu=1}^k q^{\downarrow}_{\nu}$, and equality holds for $k=N$. The superscript $\downarrow$ means that the components of the vector are arranged in descending order. Equation (\ref{Maj-un}) is universal in the sense that different entropic uncertainty relations can be obtained by applying different Schur-concave functions on it \cite{Maj-1}.

Since the work of Maccone and Pati \cite{mp}, more tightened lower bounds were developed in both product  \cite{Debasis,song} and sum forms \cite{Debasis,song,sun,Dodonov} for the variance-based uncertainty relations. The generalizations to incorporate $N$ observables \cite{xiao,chenfei,songqc} and state-independent lower bounds \cite{L2, Branciard, Guise} were also obtained. In spite of the great effort devoted to the subject, getting the optimal lower bounds of the entropic uncertainty relations remains a challenging task \cite{Entropy-app} and a hot topic \cite{Improve-entropy-un}. Contrary to the variance-based and entropic uncertainty relations, the optimal bound problem of the majorization uncertainty relations has just been solved by exploring the lattice theory \cite{Maj-latt}.

On the experimental side, the verifications of the uncertainty relations of \cite{mp} were carried out using the single photon states generated via spontaneous parametric down-conversion (SPDC) \cite{Xue-1}. Using the single spin in the nitrogen-vacancy (NV) center, an uncertainty relation for triple observables has been experimentally verified  \cite{dufei}. Systematic comparisons of typical variance-based uncertainty relations through experiment are carried out in \cite{zhixin1} with SPDC single photon source (see \cite{zhixin2, Xue-2} for latest developments). There are also the experiments of the entropic uncertainty relations using NV center \cite{NV-entropy}. In Ref.\cite{Single-Maj-exp}, the majorization uncertainty relation was experimentally investigated using the Schur-concave function. However, measuring the Schur-concave function is not sufficient for the majorization relation between distributions, e.g., we have $\vec{p} \prec \vec{q} \Rightarrow H(\vec{p}\,) > H(\vec{q}\,)$ for the Schur-concave function of Shannon entropy, but $H(\vec{p}\,) > H(\vec{q}\,) \nRightarrow \vec{p} \prec \vec{q}$ \cite{IEEE-assisted}. A faithful representation of the majorization uncertainty relation using the Lorenz curve has been introduced in \cite{Maj-latt}, where $\vec{p}\prec \vec{q}$ if and only if the Lorenz curve of $\vec{p}$ is everywhere below (enclosed by) that of $\vec{q}$.

Existing experiments about uncertainty relation mostly focus on the measurement of each individual quantum. As the uncertainty relations discussed here are ensemble properties of the quantum system (the preparation uncertainty relation), a high and stable flux of quanta would be suitable for the verification of the uncertainty relation. In this paper, we express the uncertainty relations for the two-dimensional quantum mechanical system in terms of the Stokes parameters of a polarized beam. Two typical variance-based uncertainty relations are then tested and the results are compared with the previous ones from single photon source measurement. Using the same photon source, we perform a direct and faithful experimental test of the optimal majorization uncertainty relation by measuring the Lorenz curves.

\section{The Stokes parameters}

Let $|H\rangle$ and $|V\rangle$ be the horizontal and vertical polarizations of a photon, then $45^{\circ}$ ($|+\rangle$) and $135^{\circ}$ ($|-\rangle$), left ($|L\rangle$) and right ($|R\rangle$) handed polarizations are
\begin{align}
|+\rangle& = \frac{1}{\sqrt{2}}(|H\rangle + |V\rangle) \; , \;\; |-\rangle = \frac{1}{\sqrt{2}}(|H\rangle - |V\rangle) \; , \\
|L\rangle &= \frac{1}{\sqrt{2}}(|H\rangle + i|V\rangle) \; , \; |R\rangle = \frac{1}{\sqrt{2}}(|H\rangle - i|V\rangle) \; .
\end{align}
A polarized beam forms an ensemble of a two-dimensional quantum mechanical system and the polarization state of the ensemble can be determined by the four intensity measures called Stokes parameters \cite{Book-stokes}
\begin{align}
\mathcal{S}_0 & := I_{H}  + I_{V} \; , \; \mathcal{S}_1 := I_{H} - I_{V} \; \notag, \\
\mathcal{S}_2& := I_{+} - I_{-} \; , \; \mathcal{S}_3 := I_{R} - I_{L}\;, \label{stokes}
\end{align}
where the subscripts of the intensity $I$ denote the corresponding polarizations. According to the quantum theory of photon detection, the classical intensity of radiation field is proportional to the number of photons counted by detector averaged over a time. The Stokes parameters are then defined as \cite{James}
\begin{align}
\mathcal{S}_0 & = \mathcal{N}( \langle H|\rho|H\rangle + \langle V|\rho|V\rangle ) \; , \\
\mathcal{S}_1 & = \mathcal{N}( \langle H|\rho|H\rangle - \langle V|\rho|V\rangle ) \; , \\
\mathcal{S}_2 & = \mathcal{N}(\langle H|\rho|V\rangle +\langle V|\rho|H\rangle) \; , \\
\mathcal{S}_3 & = \mathcal{N}i(\langle V|\rho|H\rangle -\langle H|\rho|V\rangle)\; .
\end{align}
Here, $\mathcal{N}$ is a constant depending on the detector efficiency and beam intensity; $\rho$ is the density matrix describing the polarization state of the ensemble. In terms of Stokes parameters, the density matrix may be expressed as
\begin{equation}
\rho = \frac{1}{2} \sum_{i=0}^3 \frac{\mathcal{S}_i}{\mathcal{S}_0} \hat{\rho}_i \; ,
\end{equation}
where $\hat{\rho}_0$ is an identity operator, $\hat{\rho}_1 = |H\rangle\langle H|- |V\rangle\langle V|$, $\hat{\rho}_2 = |H\rangle\langle V|+ |V\rangle\langle H|$, and $\hat{\rho}_3 = i ( |H\rangle\langle V|- |V\rangle\langle H|)$. When working in the $|H\rangle$ and $|V\rangle$ bases of the two-dimensional system, $\hat{\rho}_i$ is related with the Pauli operators as
\begin{align}
\sigma_x = \hat{\rho}_{2}\; , \; \sigma_y = -\hat{\rho}_{3}\; , \; \sigma_z = \hat{\rho}_{1} \; . \label{pauli-opt}
\end{align}
Then the expectation value of an arbitrary spin observable $\vec{\sigma}\cdot \vec{n}$ along unit direction $\vec{n}=(n_x,n_y,n_z)$ can be evaluated via
\begin{align}
\langle \vec{\sigma} \cdot \vec{n} \rangle = \mathrm{Tr}[\rho(\vec{\sigma} \cdot \vec{n})] = \frac{1}{\mathcal{S}_0} \widetilde{\mathcal{S}} \cdot \vec{n}\; , \label{Obs-Stokes}
\end{align}
with $\widetilde{\mathcal{S}} = (\mathcal{S}_2, -\mathcal{S}_3, \mathcal{S}_1)$.

On the other hand, according to classical theory of optical coherence, the polarization of light can be described by the coherency matrix $G$ whose elements are correlation functions. For a transverse electromagnetic wave traveling in the $z$ direction, the Stokes parameters are related to $G$ in the following form \cite{Book-stokes}
\begin{align}
\mathcal{S}_0 = G_{xx} + G_{yy} \; ,& \; \mathcal{S}_1 = G_{xx} - G_{yy} \; , \label{S-cohe-1} \\
\mathcal{S}_2 = G_{xy} +  G_{yx} \; ,& \; \mathcal{S}_3 = -i(G_{xy} -  G_{yx})\;. \label{S-cohe-2}
\end{align}
Here $G_{xx}$ and $G_{xy}$ are the autocorrelation and cross-correlation functions. The inequality $|G_{xy}|^2 \leq G_{xx}G_{yy}$ leads to the condition $\mathcal{S}_1^2+\mathcal{S}_2^2+\mathcal{S}_3^2 \leq \mathcal{S}_0^2$ \cite{Book-stokes}. Therefore, the classical partially  polarized light can also be characterized by measuring the Stokes parameters via Eq.(\ref{stokes}). Note that, the correlation functions in $G$ are of first order (see chapter 11.4 of \cite{Book-stokes}) and for light fields with identical spectral properties it is not possible to distinguish the nature of the light source from the first order correlation function, i.e., a laser beam and a conventional thermal source (see chapter 4.4 of \cite{Book-Scully}). Following we shall express the target variance-based uncertainty relations and optimal universal uncertainty relation in Stokes form for experimental preparation.

The uncertainty relation
\begin{align}
\sum_{i\in \{x,y,z\}}\Delta \sigma_i^2 \geq \frac{2}{\sqrt{3}} \left( |\langle \sigma_x\rangle| +|\langle \sigma_y\rangle| +|\langle \sigma_z \rangle| \right)  \label{our2}
\end{align}
has been verified in experiments of single-photon measurement \cite{zhixin1} and single spin in diamond \cite{dufei}. In the general case of $N$ incompatible observables encompassed, a stronger uncertainty relation writes \cite{chenfei}
\begin{align}\label{chen}
&\sum_{i=1}^{N}(\Delta A_{i})^{2}
\geq\frac{1}{N-2}\sum_{1\leq i<j\leq N}\left[\Delta (A_{i}+A_{j})\right]^2\notag\\
&-\frac{1}{(N-1)^2(N-2)}\left[\sum_{1\leq i<j\leq N}\Delta (A_{i}+A_{j})\right]^2\; .
\end{align}
This uncertainty relation was demonstrated for the $N=3$ case by using single photons through SPDC \cite{zhixin1}.

The optimal majorization uncertainty relation for $N$ observables was recently obtained \cite{Maj-latt}
\begin{align}
\vec{p}_1 \oplus \vec{p}_2\oplus \cdots \oplus \vec{p}_N \prec \vec{s} \; ,  \label{mjl}
\end{align}
and it is desirable to provide experimental verifications. Here $\vec{p}_i$ represents the probability distribution of the measurement outcomes of $A_i$; $\vec{s}$ denotes a state independent vector. For two and three observables of two-dimensional quantum mechanical system, the majorization universal uncertainty relation (\ref{mjl}) predicts
\begin{align}
& \vec{p}_z \oplus \vec{p}_x  \prec \vec{s} \;, \label{Maj-eq-xz} \\
& \vec{p}_z \oplus \vec{p}_x \oplus \vec{p}_y \prec \vec{s}\,' \; . \label{Maj-eq-xyz}
\end{align}
Here $\vec{p}_{x,y,z}$ denote those probability distributions of observables $\sigma_{x,y,z}$; the vectors $\vec{s} =(
1, \frac{\sqrt{2}}{2}, \frac{2-\sqrt{2}}{2}, 0)^{\mathrm{T}}$ and $\vec{s}\,' = (1, \frac{\sqrt{2}}{2}, \frac{1+\sqrt{3}- \sqrt{2}}{2}, \frac{1-\sqrt{3}+ \sqrt{2}}{2},  \frac{2-\sqrt{2}}{2}, 0)^{\mathrm{T}}$ \cite{Maj-latt}.

Based on equation (\ref{Obs-Stokes}), the observable quantities in the above uncertainty relations can be expressed in form of Stokes parameters. A simple substitution turns equation (\ref{our2}) to
\begin{align}
3-V \geq \frac{2}{\mathcal{S}_0\sqrt{3}}(|\mathcal{S}_1|+ |\mathcal{S}_2|+ |\mathcal{S}_3|) \; ,
\label{stokesour2}
\end{align}
with $V = (S_1^2+S_2^2+S_3^2)/S_0^2$.  For $N=3$ and $A_1=\sigma_x$, $A_2=\sigma_y$, and $A_3=\sigma_z$, equation (\ref{chen}) becomes
\begin{align}
3-V&\geq2(3-V-D)-\frac{1}{4}(L_{12}+K_{23}+ K_{13})^2\; . \label{uncertainty-VI}
\end{align}
Here $D =(\mathcal{S}_1 \mathcal{S}_2 - \mathcal{S}_2 \mathcal{S}_3 - \mathcal{S}_1 \mathcal{S}_3)/\mathcal{S}_0^2$, $L_{12}=\sqrt{2- (\mathcal{S}_1 + \mathcal{S}_2)^2/\mathcal{S}_0^2}$ and $K_{ij} =\sqrt{2- (\mathcal{S}_i - \mathcal{S}_j)^2/\mathcal{S}_0^2}$ for $i,j = 1,2,3$. In two-dimensional system, the probability distribution of measuring $\sigma_i$ may be expressed as $\vec{p}_i = \left(\frac{1+\langle \sigma_i\rangle}{2}, \frac{1 - \langle \sigma_i\rangle}{2} \right)$, $\forall i\in \{x,y,z\}$. Equations (\ref{Maj-eq-xz}, \ref{Maj-eq-xyz}) can then be formulated as
\begin{align}
& \frac{1}{2 \mathcal{S}_0} \left[
\begin{pmatrix}
\mathcal{S}_0 + \mathcal{S}_1 \\
\mathcal{S}_0 - \mathcal{S}_1
\end{pmatrix} \oplus
\begin{pmatrix}
\mathcal{S}_0 + \mathcal{S}_2 \\
\mathcal{S}_0 - \mathcal{S}_2
\end{pmatrix} \right] \prec \vec{s} \; , \label{Maj-stokes-xz} \\
& \frac{1}{2\mathcal{S}_0}
\left[ \begin{pmatrix}
\mathcal{S}_0 + \mathcal{S}_1 \\
\mathcal{S}_0 - \mathcal{S}_1
\end{pmatrix} \oplus
\begin{pmatrix}
\mathcal{S}_0 + \mathcal{S}_2 \\
\mathcal{S}_0 - \mathcal{S}_2
\end{pmatrix} \oplus
\begin{pmatrix}
\mathcal{S}_0 - \mathcal{S}_3 \\
\mathcal{S}_0 + \mathcal{S}_3
\end{pmatrix} \right] \prec \vec{s}\,'  \; . \label{Maj-stokes-xyz}
\end{align}
While equations (\ref{stokesour2}, \ref{uncertainty-VI}) are directly verifiable by measuring the quantities on both sides of the inequalities, the demonstration of equations (\ref{Maj-stokes-xz}, \ref{Maj-stokes-xyz}) is subtle. It has been shown that the majorization relation could be verified by measuring the Lorenz curves of the probability distributions \cite{Maj-latt}. The Lorenz curve of a distribution $\vec{p}$ exhibits the function $f_{p}(n):=\sum_{\mu=1}^n p^{\downarrow}_{\mu}$, and $\vec{q} \prec \vec{p}$ if and only if the Lorenz curve of $\vec{p}$ encloses that of $\vec{q}$. Note, two vectors are incomparable under majorization relation, i.e., $\vec{p} \nprec \vec{q}$ and $\vec{p} \nsucc \vec{q}$, if and only if their Lorenz curves intersect \cite{Maj-latt}.

\section{Experimental implementation}

\begin{figure} \centering
\includegraphics[width=0.43\textwidth]{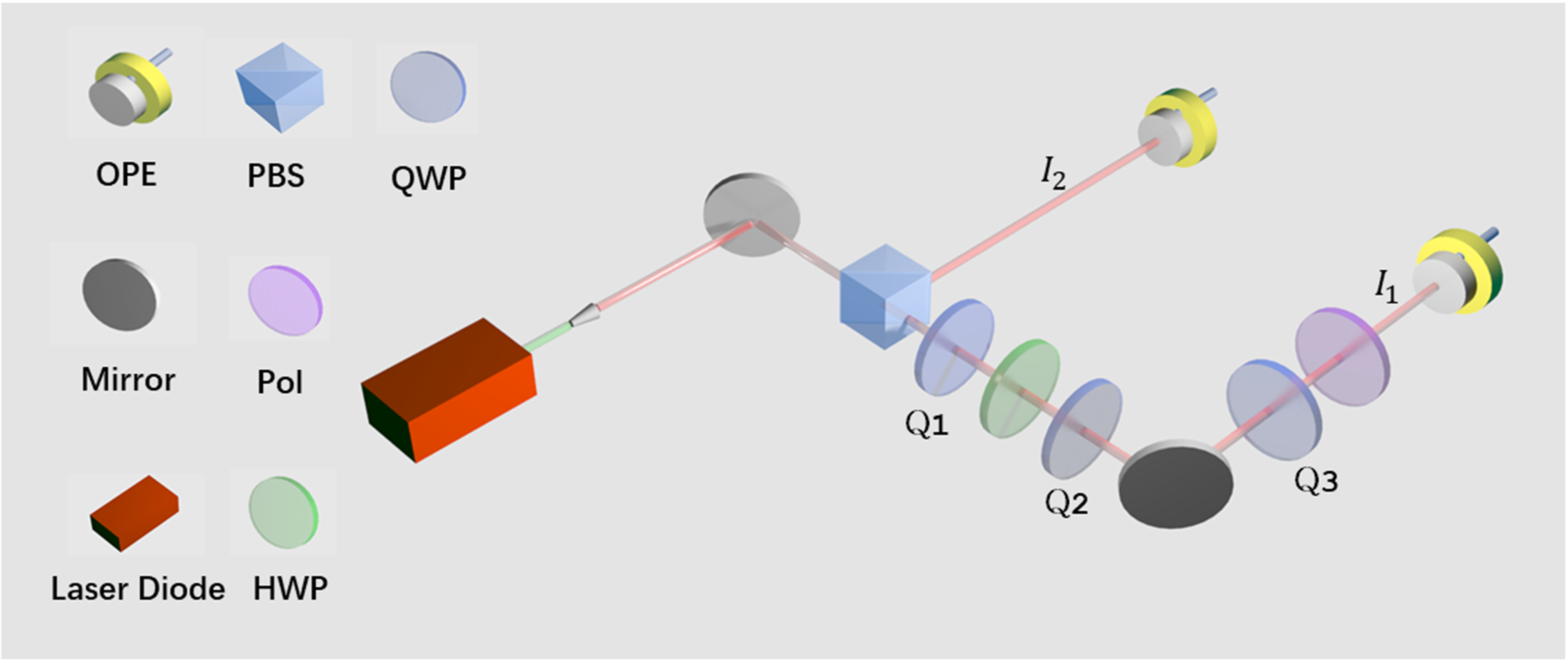}
\caption{Experimental setup. A beam passes through a polarizing beam splitter (PBS), providing linearly polarized light. The horizontal light via quarter wave plate Q1, HWP and Q2 is used to generate the two-dimensional quantum states. In the stage of measurement, the optional Q3 and the adjustable polarizer (Pol) is applied in the direction of the beam for measuring the Stokes parameters.}\label{f1}
\end{figure}

In terms of Stokes parameters $\mathcal{S}_i$, the uncertainty relations for the two-dimensional quantum mechanical system now can be experimentally verified by measuring the corresponding intensities of a polarized beam based on equation (\ref{Obs-Stokes}). Figure \ref{f1} shows the experimental setups for the preparation and measurements of the Stokes parameters of a polarized beam. A  6-mW  tested beam originated from a fiber-coupled diode laser with a wavelength of 808nm. The power stability is 0.5(\%) in long duration. In the stage of preparation, the laser passes through a polarizing beam splitter (PBS), providing linearly polarized light. The transmission path is used to prepare the required state of the light, and the intensity $I_2$ in another arm is used as a reference. The intensities are directly detected by the optical power meter (OPE). The resolution of OPE is 100 pW within the response time of 1 $\upmu$s, giving sufficient recognition to measure the variation of light intensity.

\begin{figure}[t] \centering
\scalebox{0.31}{\includegraphics{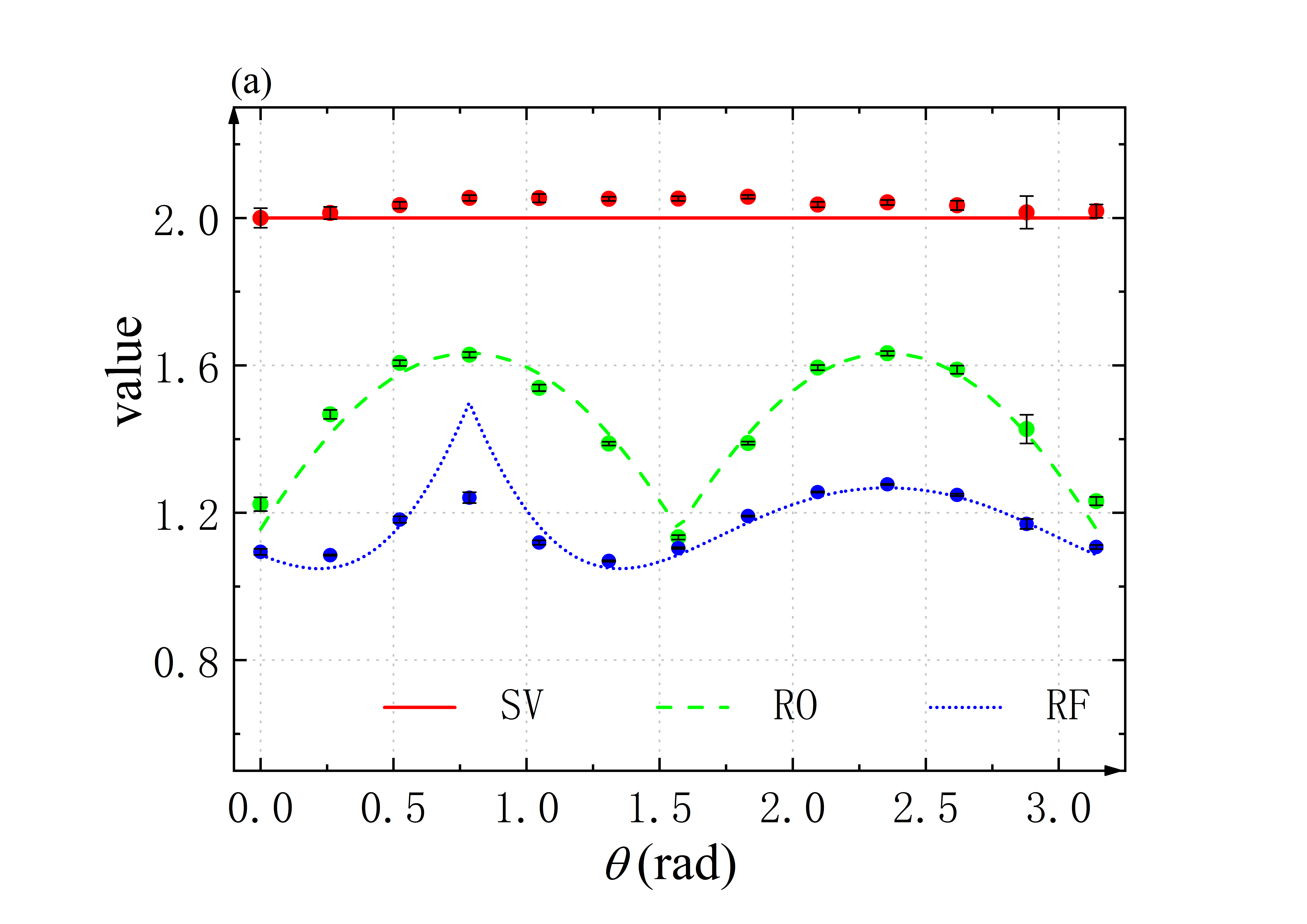}}
\scalebox{0.31}{\includegraphics{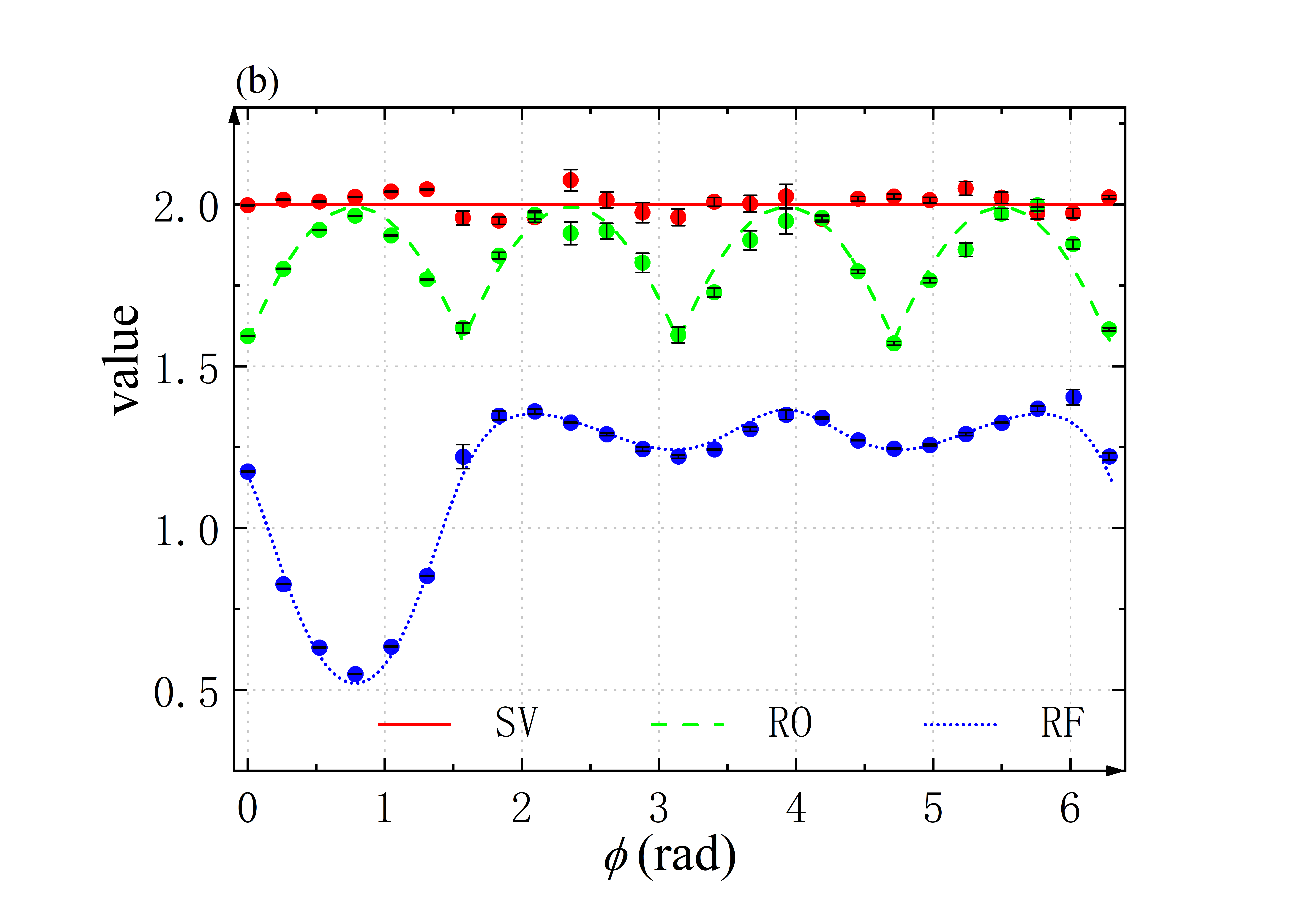}}
\caption{Experimental results for variance-based uncertainty relations. ({\bf a}) Experimental results with states $|\psi(\theta,0)\rangle$. ({\bf b}) Experimental results with states $|\psi(\pi/3,\phi)\rangle$. The solid red line corresponds to the LHS of inequalities (\ref{uncertainty-VI}) and (\ref{stokesour2}), i.e. the sum of variances (SV) $(\Delta \sigma_{x})^{2}+(\Delta \sigma_{y})^{2}+(\Delta \sigma_{z})^{2}$. The red circles represent the measured sum of variances. The dotted blue and dashed green curves represent the theoretical values of RF and RO, where RF and RO denote the RHS of relations (\ref{uncertainty-VI}) and (\ref{stokesour2}) respectively. The blue and green circles, in turn, represent the experimental values of RF and RO. Error bars indicate $\pm1$ standard deviation.
}\label{f2}
\end{figure}

On the path of transmission, the horizontal light via quarter wave plate (QWP) Q1, half-wave plate (HWP) and QWP Q2 is set to an arbitrary state of the Bloch sphere $|\psi(\theta,\phi)\rangle=\cos\frac{\theta}{2}|H\rangle+e^{i \phi}\sin \frac{\theta}{2} |V\rangle$. In the stage of measurement, the optional Q3 and the adjustable polarizer (Pol) is applied in the direction of the beam for measuring the Stokes parameters. Finally, the OPE records the intensity distribution $I_1$ of different projection directions. Note that in order to achieve better experimental results, it is inevitable to measure noise or consider the rate of signal to reference $I_1/I_2$ as the real intensity, which can eliminate the impact of the beam energy fluctuation and improve the quality of polarization measurement.

To test inequalities (\ref{stokesour2}-\ref{Maj-stokes-xyz}), we choose a series of states $|\psi(\theta,0)\rangle$ and $|\psi(\frac{\pi}{3},\phi)\rangle$, where $\theta$ or $\phi = n \pi / 12$. Those system states can be generated by tuning the setting angles of Q1 and HWP, and Q2 is set to be $45^{\circ}.$ The experimental results of inequalities are calculated from the measurement of relative Stokes parameters ${S_i}/{S_0}$. We demonstrate uncertainty relations (\ref{stokesour2}, \ref{uncertainty-VI}) in Figure \ref{f2}. All experimental data fit the theoretical predictions well and the bound (\ref{stokesour2}) are above the curves of (\ref{uncertainty-VI}) when $N=3$. The results indicate that uncertainty relation (\ref{our2}) is more tighter for three incompatible observables. The fidelity of the experimental result is more than 98.5\%, which is calculated from
$F \left( \rho, \rho _ { 1 } \right) = \operatorname { tr } \left( \rho \rho _ { 1 } \right) + \sqrt { 1 - \operatorname { Tr } \left( \rho^ { 2 } \right) } \sqrt { 1 - \operatorname { Tr } \left( \rho _ { 1 } ^ { 2 } \right) },$
with
$\rho _ { 1 }=|\psi(\theta,\phi)\rangle \langle\psi(\theta,\phi)|.$

\begin{figure}[t] \centering
\scalebox{0.31}{\includegraphics{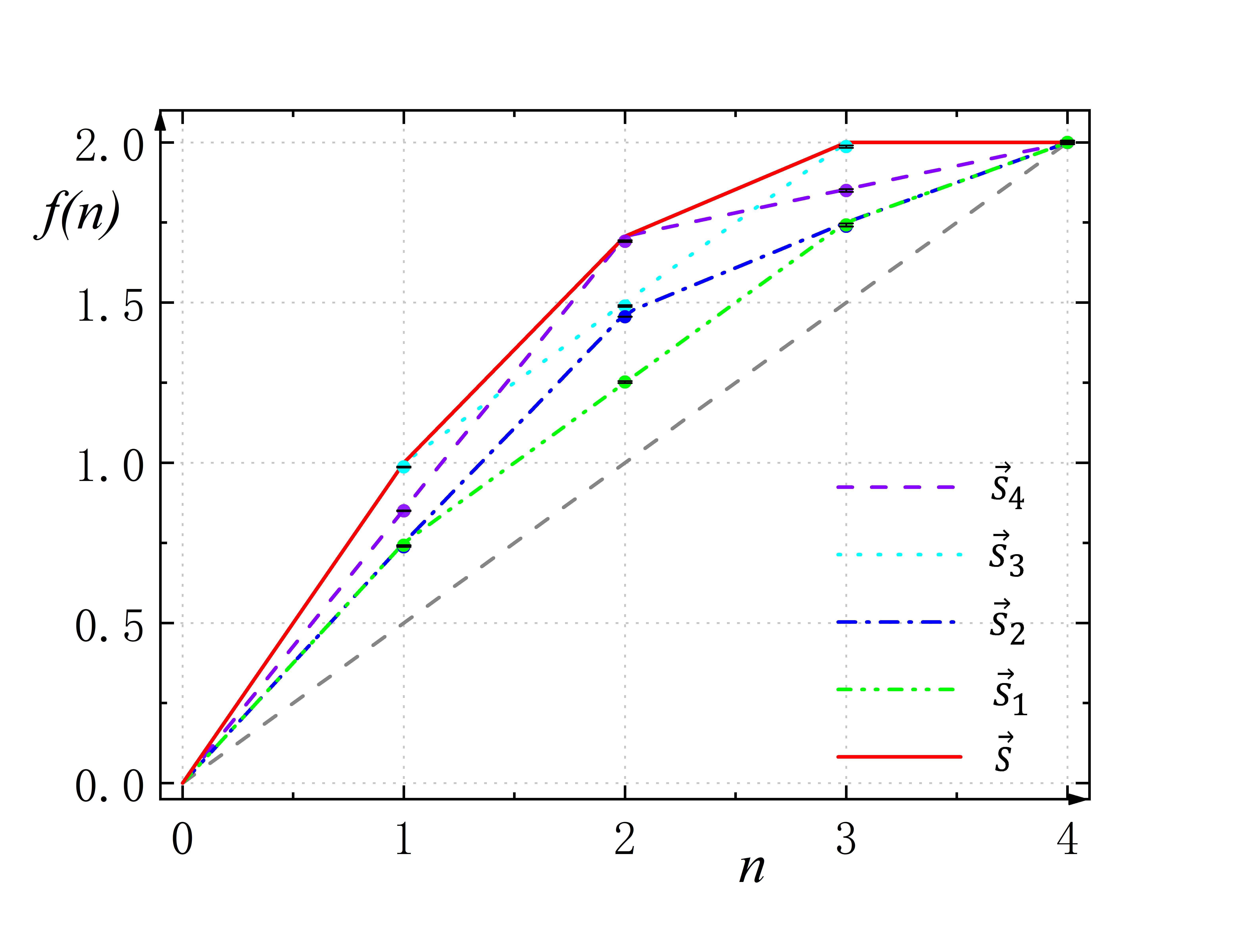}}
\caption{Experimental results for the direct-sum majorization uncertainty relation with observables $\sigma_{z}$ and $\sigma_{x}$. The line plots of the discrete functions $f (n)$ forms the Lorenz curves of the distributions. The red solid line is the Lorenz curve of $\vec{s}$ on the RHS of inequality (\ref{Maj-stokes-xz}). The doubly dotted green, dot-dashed blue, dotted
cyan, and dashed violet curves represent the theoretical predictions for the Lorenz curves of $\vec{s}_{1},\vec{s}_{2},\vec{s}_{3}$ and $\vec{s}_{4}$ which are given by the states $|\psi(\frac{\pi}{3}, \frac{\pi}{2})\rangle$, $|\psi(\frac{\pi}{3}, \frac{\pi}{3})\rangle$, $|\psi(\frac{\pi}{2},0)\rangle$,
and $|\psi(\frac{3\pi}{4},0)\rangle$, respectively. The corresponding colored circles represent the experimental values of  $\vec{s}_{1},\vec{s}_{2},\vec{s}_{3}$ and $\vec{s}_{4}$ by turns. Error bars represent $\pm1$ standard deviation.
}\label{f4}
\end{figure}

In Figure \ref{f4}, we show the experimental results of the universal uncertainty relation (\ref{Maj-stokes-xz}) by depicting the distributions of the left and right hand sides of ``$\prec$" in terms of Lorenz curves. The red solid line in Figure \ref{f4} represents the Lorenz curve of $\vec{s}$. In the experiment, we choose four states $|\psi(\frac{\pi}{3}$, $\frac{\pi}{2})\rangle$, $|\psi(\frac{\pi}{3}$, $\frac{\pi}{3})\rangle$, $|\psi(\frac{\pi}{2},0)\rangle$,
and $|\psi(\frac{3\pi}{4},0)\rangle$, which corresponds to the Lorenz curves of $\vec{s}_{1},\vec{s}_{2},\vec{s}_{3}$ and $\vec{s}_{4}$, as shown in Figure \ref{f4}. The experimental data for the LHS of relation (\ref{Maj-stokes-xz}), i.e. $\vec{s}_i$, are obtained by the measuring the Stokes parameters. The result shows that the corresponding Lorenz curves of $\vec{s}_i$ all lie below that of $\vec{s}$, say enclosed by Lorenz curve of $\vec{s}$. That means $\vec{s}_i \prec \vec{s}$ is satisfied for all $i=1,2,3,4$. It should be noted that though experimental results show $H(\vec{s}_4)=1.24 > H(\vec{s}_3) =1.102 $, one cannot conclude $\vec{s}_4 \prec \vec{s}_3$ (they have intersected Lorenz curves). This is a peculiar characteristics of majorization relation, which is detectable only through the Lorenz curve measurement.

\begin{figure}[t] \centering
\scalebox{0.31}{\includegraphics{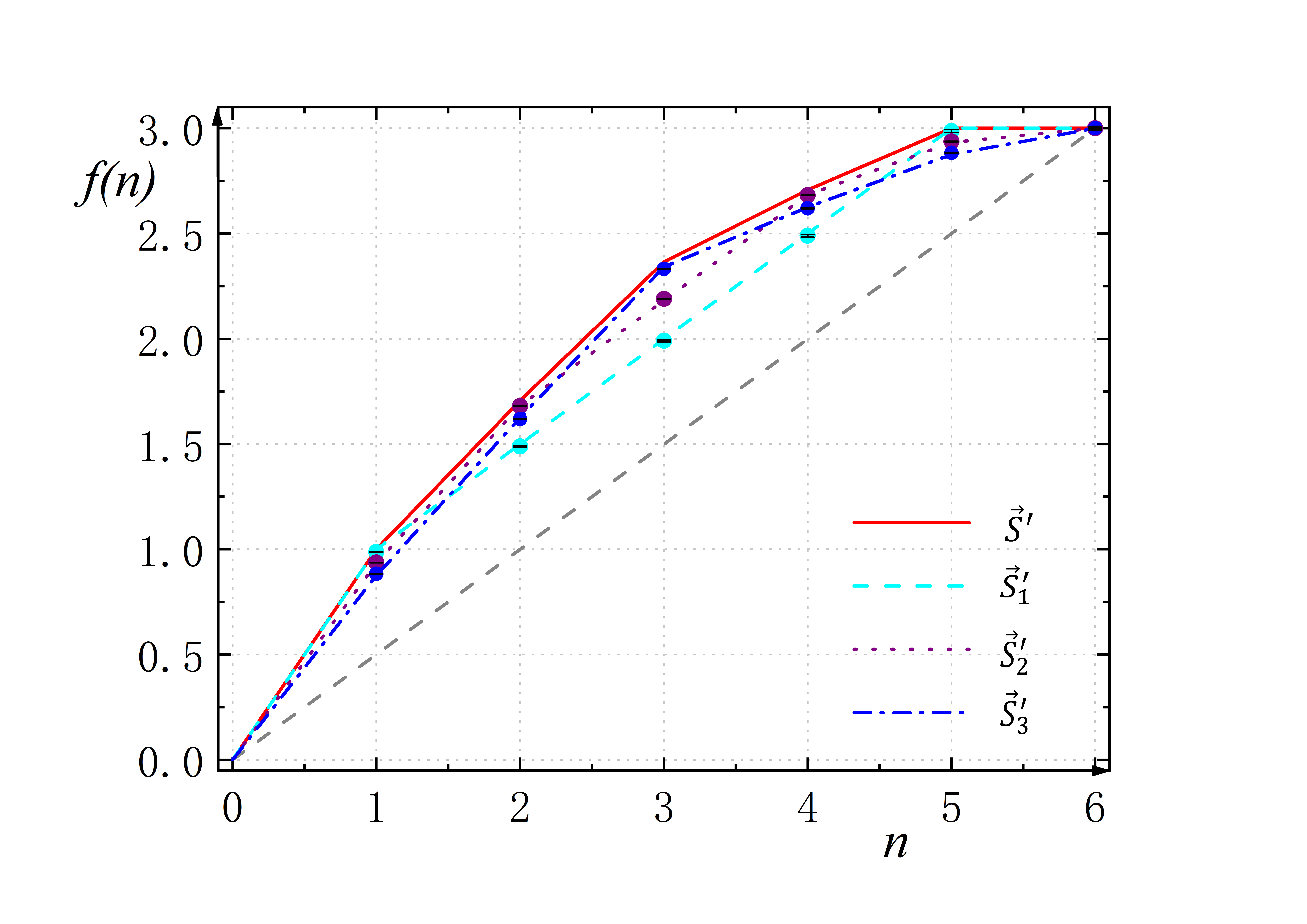}}
\caption{Experimental results for the direct-sum majorization uncertainty relation with observables $\sigma_{z}$, $\sigma_{x}$, and $\sigma_{y}$. The Lorenz curves are obtained by the line plots of the discrete functions $f(n)$. The red solid line is the Lorenz curve for $\vec{s}\,'$ on the RHS of inequality (\ref{Maj-stokes-xyz}). The dashed cyan, dotted violet, and dot-dashed blue curves represent the theoretical predictions for the Lorenz curves of $\vec{s}\,'_{1},\vec{s}\,'_{2}$ and $\vec{s}\,'_{3}$ for states $|\psi(\frac{\pi}{2},0)\rangle $, $|\psi(\frac{\pi}{3},0)\rangle$
and $|\psi(\frac{\pi}{3},\frac{\pi}{6})\rangle$, respectively. The corresponding colored circles represent the experimental value of  $\vec{s}\,'_{1},\vec{s}\,'_{2}$ and $\vec{s}\,'_{3}$ by turns. Error bars represent $\pm1$ standard deviation.
}\label{f5}
\end{figure}

Figure \ref{f5} shows the experiment results for the direct-sum majorization uncertainty relation (\ref{Maj-stokes-xyz}). The dashed cyan, dotted violet, and dot-dashed blue curves represent the Lorenz curves of the left hand side of equation (\ref{Maj-stokes-xyz}) for the states of $|\psi(\frac{\pi}{2},0)\rangle $, $|\psi(\frac{\pi}{3},0)\rangle$ and $|\psi(\frac{\pi}{3},\frac{\pi}{6})\rangle$, labeled by $\vec{s}\,'_{\!1}$, $\vec{s}\,'_{\!2}$ and $\vec{s}\,'_{\!3}$, respectively, as shown in Figure \ref{f5}. All of them are enclosed by the Lorenz curve of $\vec{s}\,'$, the red solid line in Figure \ref{f5}. Hence, $\vec{s}\,'_{\!i} \prec \vec{s}\,'$ is satisfied for all $i=1,2,3$.

Our results on the variance-based uncertainty relations of equations (\ref{our2}-\ref{chen}) agrees with the predictions of quantum mechanics, and also with previous measurements using single photon sources, e.g. see Figures 2 and 3 in Ref. \cite{zhixin1}. The direct-sum majorization uncertainty relations (\ref{Maj-eq-xz}, \ref{Maj-eq-xyz}) are also verified here. Existing experiments using SPDC photons or the single spin state in NV center can be regarded as using the single photon number state (or Fock state). Our experiment uses the laser field, i.e., the coherent state. Considering the discussions after the equations (\ref{S-cohe-1}) and (\ref{S-cohe-2}), the thermal light is supposed to be capable of carrying out the experiments of measuring the uncertainty relations either. Therefore, it may be expected that different statistics of the ensemble of quanta, i.e., the Fock state, coherent state, or thermal state, should not affect the preparation of uncertainty relation for each individual particle.

The optimality of the universal uncertainty relation is propped by the fact that some experimental data points on the Lorenz curves of the quantum states coincide with that of $\vec{s}$ and $\vec{s}\,'$. In other words, the bound reaches to its minimum. In our experiment, the error bars induced by the measure of intensity are rather small, because it is less affected by noises comparing to the case of single photon source. The power stability of our photon source is about $10^{-3}$(0.5\%), and a high stability of $10^{-4}$ has been reported in literature \cite{power-yang}. Although the system state has high fidelity, there are still data points drift from the theory predictions in our experiment. The main sources of this experimental error may attribute to the imperfectness of calibration and the retardation of wave plate.

\section{Conclusions}

In this work, we experimentally verified two different types of uncertainty relations, the variance-based uncertainty relations and the optimal universal uncertainty relation, in two-dimensional quantum mechanical system by exploring the intensity measures of a coherent beam. It is achieved by relating the expectation values of the observables with the Stokes parameters of the polarized beam. Of the variance-based uncertainty relation, we confirm previous measurements using single photon source, but with a high precision. More importantly, we have made the first direct experimental test of the optimal majorization universal uncertainty relation by measuring the Lorenz curves of the majorization relations. The incomparability of two distributions under the majorization is exhibited in our experiment. Since the measure precision of intensity relies on the laser power stability, which has now reached $10^{-4}$ \cite{power-yang}, the scheme of Stokes parameters measurement makes the ultrahigh precision test of uncertainty relation to be reachable.

\section*{Acknowledgements}
\noindent
This work was supported in part by the Ministry of Science and Technology of the Peoples' Republic of China(2015CB856703); by the Strategic Priority Research Program of the Chinese Academy of Sciences, Grant No.XDB23030100; by the National Natural Science Foundation of China(NSFC) under the Grants 11375200 and 11635009; and by the University of Chinese Academy of Sciences.

\vspace{1cm}
\section*{Appendix}

We have verified the uncertainty relations by measuring Stokes parameters of the radiation field with typical polarization states. Here we supplement the experimental verification of the majorization uncertainty relations using more states. We can prepare an arbitrary state with a PBS, several setting angles wave plates. For example, a state $|\psi(\pi/3,\pi/6)\rangle$ can be realized when the setting angles of Q1, HWP and Q2 are $15 ^ { \circ }, 37.5 ^ { \circ }$ and $45 ^ { \circ } $, respectively. Then, project into the directions of $|H\rangle$ ,$|V\rangle$ , $|+\rangle$, $|-\rangle$ ,$|L\rangle$ and $|R\rangle$, after applying the Stokes parameters $\mathcal{S}_i$, we obtain the probability $\vec{p}_{x,y,z}$. This method directly demonstrate the optimal universal uncertainty relation from the ensemble property.

\begin{figure}\centering
\scalebox{0.31}{\includegraphics{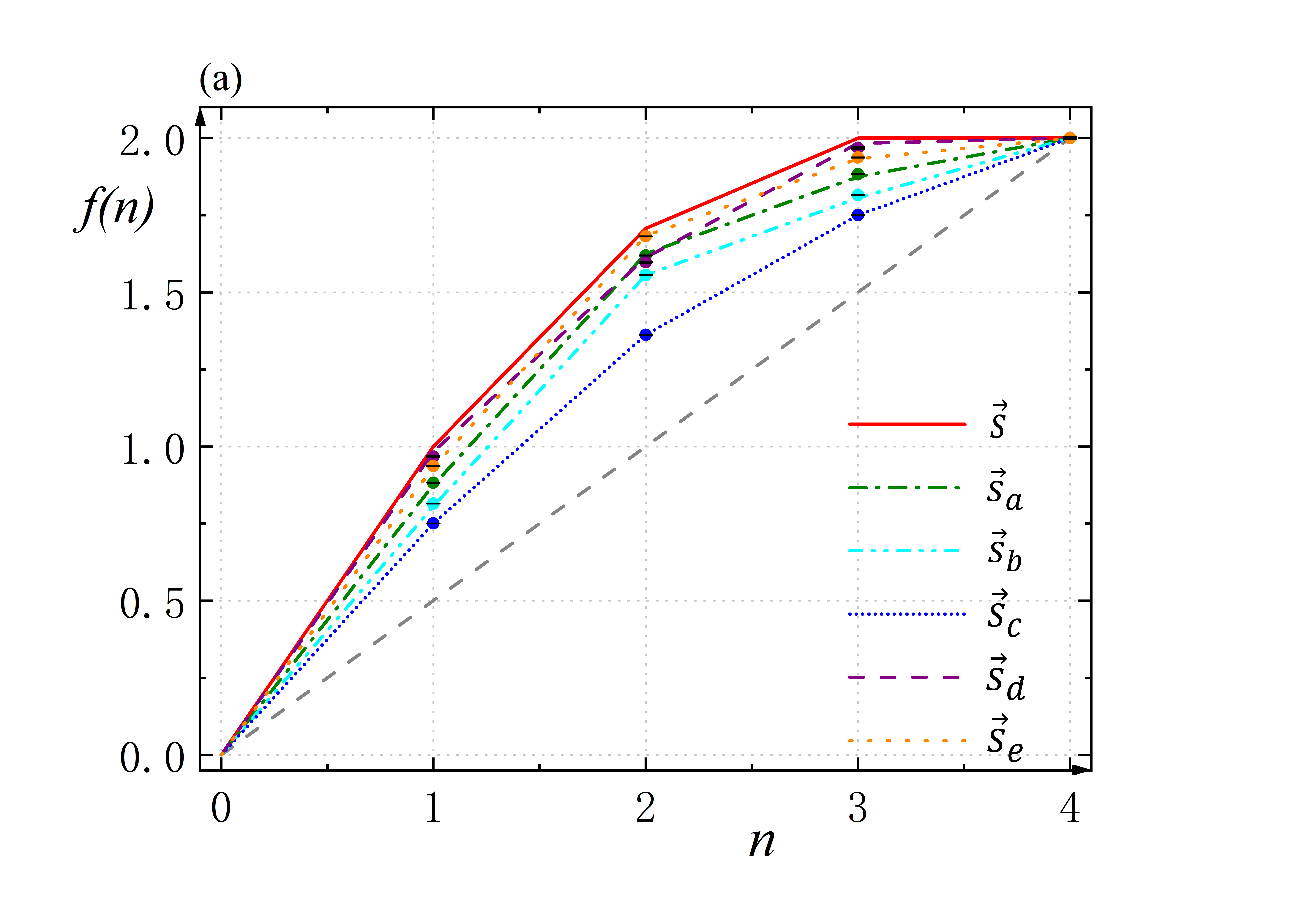}}
\scalebox{0.31}{\includegraphics{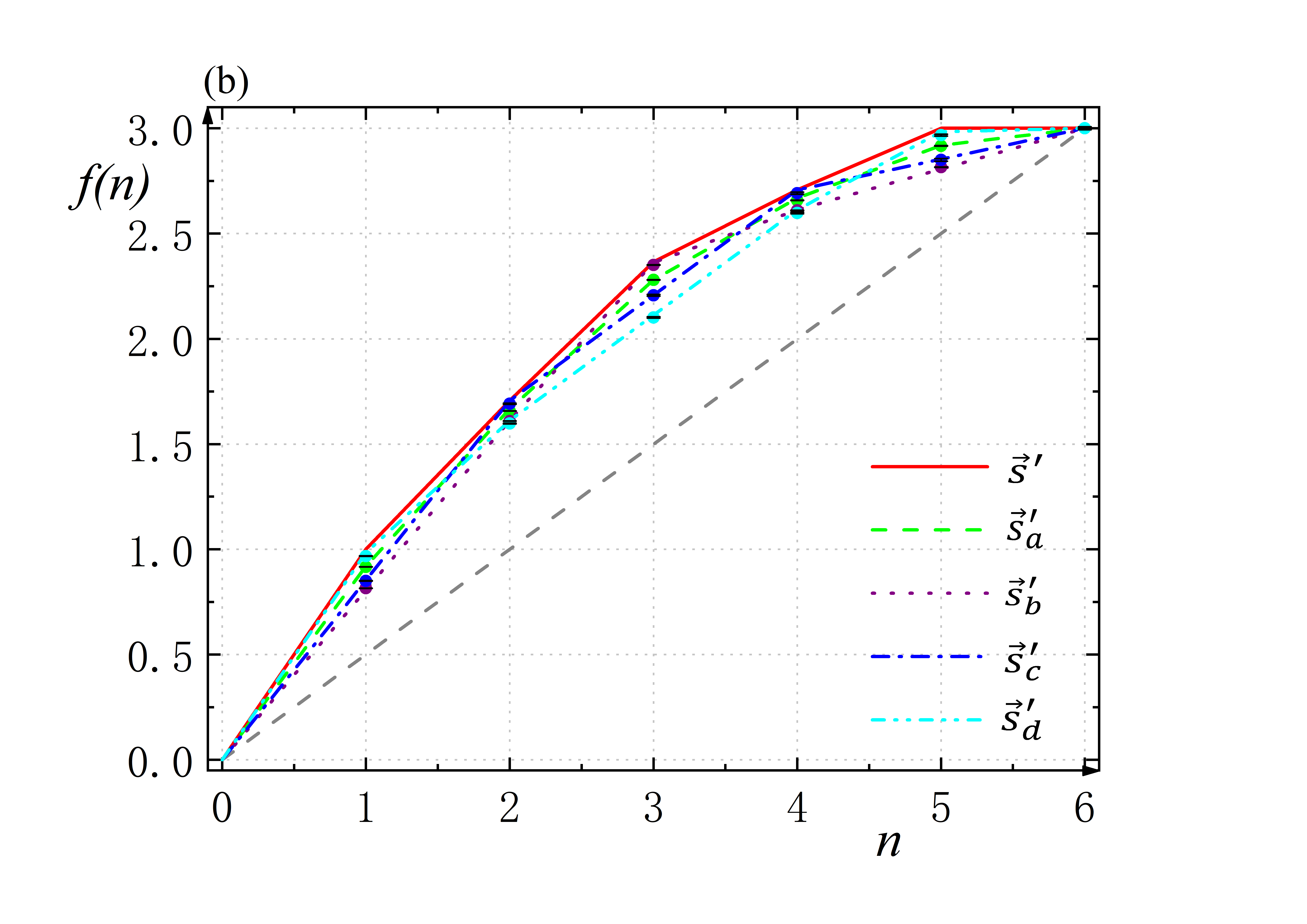}}
\caption{Experimental results for the direct-sum majorization uncertainty relations. ({\bf a}) Supplemental demonstration of direct-sum majorization uncertainty relation for the observables $\sigma_{z}$ and $\sigma_{x}$ with states $|\psi(7\pi/12,0)\rangle$ and $|\psi(\pi/3,n\pi/12)\rangle$ for $n=0,2,3,5$, whose Lorenz curves we denoted with $\vec{s}_{a}-\vec{s}_{e}$, respectively.
({\bf b})  Supplemental demonstration of direct-sum majorization uncertainty relation for the observables $\sigma_{z}$, $\sigma_{x}$, and $\sigma_{y}$ with states $|\psi(\pi/3,\pi/12)\rangle$, $|\psi(\pi/3,\pi/4)\rangle$, $|\psi(3\pi/4,0)\rangle$ and $|\psi(7\pi/12,0)\rangle$, whose Lorenz curves we denoted with $\vec{s}\,'_{a}-\vec{s}\,'_{d}$. In both ({\bf a}) and ({\bf b}), the red curves of $\vec{s}$ and $\vec{s}\,'$ corresponds to the bound of inequalities  (\ref{Maj-stokes-xz}, \ref{Maj-stokes-xyz})  and the Lorenz curve of the prepared states all lie below that.  Error bars represent $\pm1$ standard deviation.
}\label{f6}
\end{figure}

We noticed that the probability distributions of the observables $\vec{p}_{x,y,z}$ may be the same in different system states. In Figure \ref{f6}, we show the supplemental experimental results for existed various probability distributions. We choose $|\psi(\pi/3,0)\rangle$, $|\psi(\pi/3,\pi/6)\rangle$, $|\psi(\pi/3,\pi/4)\rangle$, $|\psi(\pi/3,5\pi/12)\rangle$ and $|\psi(7\pi/12,0)\rangle$ in Figure \ref{f6}(a). The Lorenz curves of these prepared states are enclosed by the red curves (the bound of inequalities), and so is in Figure \ref{f6}(b). Thus our setup
completed the  investigation of the optimal majorization universal uncertainty relation in two-dimensional quantum mechanical system.

\end{document}